\documentclass[conference]{IEEEtran}
\IEEEoverridecommandlockouts
\usepackage{cite}
\usepackage{subcaption}
\usepackage[hidelinks]{hyperref}
\usepackage{amsmath,amssymb,amsfonts}
\usepackage{algorithmic}
\usepackage{graphicx}
\usepackage{textcomp}
\usepackage{xcolor}
\usepackage{mathtools}
\usepackage{booktabs}
\usepackage{siunitx}
\usepackage[export]{adjustbox}
\usepackage{balance}

\begin{document}

\title{Analysis of Distributed Congestion Control in Cellular Vehicle-to-everything Networks}

\author{Behrad Toghi$^*$, Md Saifuddin$^*$, Yaser P. Fallah$^*$, M. O. Mughal$^*$\\
$^*$Connected and Autonomous Vehicle Research Lab (CAVREL)\\
University of Central Florida, Orlando, FL\\
\{toghi, md.saif, ozair\}@knights.ucf.edu, yaser.fallah@ucf.edu
        }
\maketitle
\begin{abstract}
Cellular Vehicle-to-everything (C-V2X) communication has been proposed in the 3rd Generation Partnership Project release 14 standard to address the latency and reliability requirements of cooperative safety applications. Such applications can involve highly congested vehicular scenarios where the network experiences high data loads. Thus, a sophisticated congestion control solution is vital in order to maintain the network performance required for safety-related applications. With the aid of our high-fidelity link-level network simulator, we investigate the feasibility of implementing the distributed congestion control algorithm specified in SAE J2945/1 standard on top of the C-V2X stack. We describe our implementation and evaluate the performance of transmission rate and range control mechanisms using relevant metrics. Additionally, we identify areas for potential design enhancements and further investigation.
 \end{abstract}
\begin{IEEEkeywords}
congestion control, scalability, LTE  mode-4, LTE-V, semi-persistent scheduling (SPS), cellular vehicle-to-everything (C-V2X)
\end{IEEEkeywords}
%
\section{Introduction}
Cooperative vehicle safety (CVS) applications are introduced to enhance safety end efficiency of intelligent transportation systems \cite{hnmahjoub:vtc, hnmahjoub:cavs, hnmahjoub:syscon}. Dedicated Short Range Communication (DSRC) designed in 1999 as the primary vehicular communication solution and was dominant for years before the introduction of Cellular Vehicle-to-everything (C-V2X) by 3rd Generation Partnership Project (3GPP).
Support for vehicle-to-everything (V2X) communication was added
in order to the 3GPP release 14 standard
to satisfy the latency and reliability requirements needed for vehicular safety use cases.
Our focus in this work will be on C-V2X Mode-4 communication, in which user equipments (UEs) are able to communicate out of network coverage in an ad-hoc fashion. In the absence of the centralized network scheduler, UEs allocate their radio resources employing a distributed resource allocation mechanism which will be explored in the next section \cite{jgozalvez:vtm}.

In a CVS system, vehicles are required to periodically transmit their state information to increase situational awareness in the region. It is well-known that performance of such systems degrades in high density environments due to the heavy message load resulting in saturation of the wireless medium. In order to address the aforementioned concern, and thereby promote safety, congestion control algorithms have been proposed for V2X applications in the literature. We refer to \cite{yfallah:tvtsupra2016}, where the authors propose a piece-wise linear power control scheme based on channel occupancy. Authors in \cite{gbansal:limerictvt} design a linear adaptive message rate control algorithm, LIMERIC, for DSRC communication. Incorporating the insights gained from these studies, the Distributed Congestion Control (DCC) algorithm that has been standardized by the Society of Automotive Engineers (SAE) \cite{sae:j2945} for V2V networks is based on a combination of rate and power control. It is worthwhile noting that the congestion related research studies were focused on the DSRC protocol stack as it was the only V2V technology under consideration at the time. It is therefore, unclear if the resulting DCC algorithm in \cite{sae:j2945} is a technology-agnostic solution and if it would be applicable to an alternative technology that has since become available, namely C-V2X \cite{3gpp:RAN:QC:DCC}.  

Study on congestion control solutions for C-V2X in the literature is limited to one recent article \cite{bkang:cv2xPowerControl} in which authors investigate the effect of radiation power and suggest the baseline (No congestion control) performance can be maintained with lower radiation power levels. To the authors’ knowledge, there is no study on rate and range control for C-V2X in the literature. Although 3GPP Release 14 standard has specified a general congestion control framework for C-V2X \cite{3gpp:RAN:QC:DCC}, it has not provided a specific algorithm and details required for implementation in practice and the authors believe this will be up to the SAE C-V2X committee to standardize a congestion control scheme for the industry. The main contribution of this work is to assess whether the DCC algorithm can be ported onto the C-V2X stack given the fundamental differences between DSRC and C-V2X at the lower layers. We propose a hybrid design for the DCC-enabled C-V2X and evaluate its performance in dense vehicular scenarios under heavy network load, employing our network simulator \cite{btoghi:vnc}. We analyze the observations and highlight some important challenges and effects that result from the porting exercise. Based on the study, we share our conclusions on the applicability of DCC for C-V2X technology and identify directions for future study.

The rest of this text is organized as follows. In Section II, we present a tutorial-like overview of the C-V2X technology, as the underlying communication technology, in addition to DCC algorithm. Section III is dedicated to implementation details and simulation setup. In Section IV, we discuss performance of DCC in dense vehicular scenarios as well as results of parameter tuning, before concluding the paper in Section V.
\section{Background}
As mentioned earlier, DSRC and C-V2X operate very differently at the physical (PHY) and medium access (MAC) layers. Detailed discussions and analyses of C-V2X technology exist in \cite{jgozalvez:vtm, btoghi:vnc} which we encourage the reader to refer to. However, for the sake of completeness, we provide a brief overview of the C-V2X communication in this section, followed by details of DCC algorithm.
\subsection{C-V2X Mode-4 Communication}
\begin{figure}[b]
\centering
\includegraphics[width=.48\textwidth]{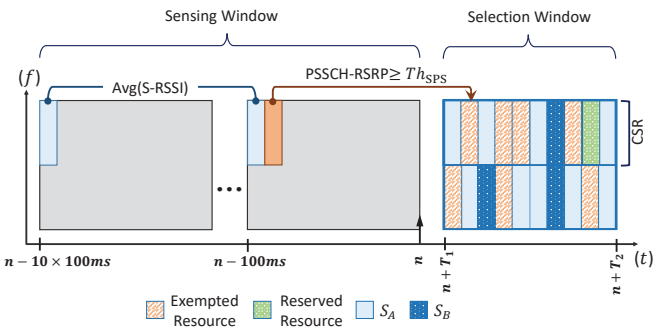}
\caption{An abstract illustration of C-V2X radio resources and its SB-SPS.}
\label{fig:fig1}
\end{figure}
Similar to traditional LTE technology, C-V2X communication employs the single-carrier frequency-division multiple access with radio resources known as Resource Blocks (RBs). Time and frequency divisions are also similar to those of the LTE technology \cite{3gpp:36213}, \emph{subframes} in time domain and \textit{subchannels} in frequency domain. Depending on the modulation and coding scheme (MCS), a group of RBs form a candidate single-subframe resource (CSR), hereafter referred to as a \emph{radio resource}. Fig. \ref{fig:fig1} illustrates the discussed time-frequency divisions in which the channel bandwidth is divided into two radio resources. The Mode-4 communication employs a distributed resource allocation scheme, known as Sensing-based Semi-persistent Scheduling (SB-SPS) \cite{3gpp:36213}. In SB-SPS mechanism, every UE senses the channel and keeps the track of received signals from its neighboring nodes during the last $1000 ms$ (sensing window). For a packet generated at time $n$, the UE utilizes information from last sensed window to schedule a set of transmission opportunities in the future.

The process of choosing a radio resource as the transmission opportunity is as follows. The UE takes a set, $S_A$, of so-called candidate radio resources in $[n+T_1, n+T_2]$ (selection window) and checks their previous occurrences in the sensing window. More precisely, if the UE has successfully received a packet which its reference signal received power (PSSCH-RSRP) is greater than a preset threshold ($Th_{\text{SPS}}$), the radio resource corresponding to that packet will be exempted from the selection window. After the exemption process, UE ranks the remainder of radio resources by their average received signal strength indicator (S-RSSI) in a descending order and selects the bottom 20\% of the list, $S_B$. If the total number of remaining resources after the exemption procedure is less than 20\% of the initial set, $S_A$, UE increments $Th_{\text{SPS}}$ by 3 dB. After choosing a transmission opportunity, the UE periodically repeats the selection for a random number of transmissions. This random number, known as sidelink resource reselection counter (SLRRC), is uniformly chosen in a predefined range. A UE changes its reservation with probability $\mathcal{P}_{\text{resel}}$ when the counter times-out \cite{3gpp:36331}.
%
\subsection{Distributed Congestion Control}
From an application layer point-of-view, information dissemination in a general CVS system can be mainly governed by two parameters, namely message broadcast rate, $r$, and range of broadcast messages, $d$. One may argue that other parameters, such as MCS, can also be utilized in the congestion control algorithm, nonetheless studies in the literature have shown that the major role is being played by two mentioned parameters, $d$ and $r$ \cite{yfallah:tvtsupra2016}. The main purpose of DCC is to mitigate excessive packet drops when a VANET is experiencing persistent packet collisions under high network load, which can lead to a totally ineffective network. This issue can be avoided by either limiting the communication range of UEs, i.e., lessening the radiated power, or by reducing their message broadcast rate.
\subsubsection{Transmission Rate Control}
As specified in \cite{sae:j2735}, UEs transmit their situational awareness messages with 10 Hz rate in the baseline operation mode. This works well as long as the channel utilization is not high. However, transmitting messages at this rate in high-density networks may exceed network's capacity and consequently degrade CVS system's performance. In order to mitigate such performance drop, rate control algorithms are proposed in the literature \cite{gbansal:limerictvt, yfallah:tits2011}. Rate control algorithm determines the inter-transmit time (ITT) based on the measured vehicle density in its proximity, $N_{STA}$. This metric is then smoothed by a single-step memory and 1/2 smoothing factor, for the stability purposes. ITT is determined using the smoothed vehicle density in range, $N_{STA}^{S}$, as shown in Equ. \ref{equ:SAEITT}.
\begin{equation} \label{equ:SAEITT}
\text{ITT} = \begin{cases}
100 \,\text{ms}, & N_{STA}^{S} \le \mathcal{B}\\ 
\frac{N_{STA}^{S}}{\mathcal{B}} \times 100 \, \text{ms},  & \mathcal{B} < N_{STA}^{S} <\frac{\text{ITT}^{\text{max}}}{100\text{ms}}\mathcal{B}\\
\text{ITT}^{\text{max}}, & \frac{\text{ITT}^{\text{max}}}{100\text{ms}}\mathcal{B}\le N_{STA}^{S}.
\end{cases}
\end{equation}
where, $\mathcal{B}$ is Density Coefficient and $\text{ITT}^{\text{max}}$ is the maximum allowed inter-transmission time.

CVS system periodically transmits messages based on the calculated ITT from Equ. \ref{equ:SAEITT}. Furthermore, since the ultimate purpose of congestion control in VANETs is maintaining the safety requirements, DCC needs a feedback from the upper-layer safety application as well. Most of CVS applications rely on positional tracking and creating a local map in a given vehicle based on the estimated location of its neighboring vehicles. Communication latency and packet drops can lead to inaccurate position estimation. Position Tracking Error (PTE) estimates the uncertainty in a vehicle's position and is defined as great-circle distance between the host-vehicle's estimated position, as perceived by remote-vehicles, and host-vehicle's actual position. It should be noted that PTE is due to the communication latency with which vehicles receive each other's kinematic state data and is distinct from positioning inaccuracies such as GPS error. In parallel to the algorithm mentioned in Equ. \ref{equ:SAEITT}, every vehicle monitors PTE and broadcasts a message if $\text{PTE}>50cm$, due to change in vehicle dynamics and inaccurate position estimation, regardless of ITT. 
\subsubsection{Transmission Range Control}
The Stateful Utilization-based Power Adaptation (SUPRA) \cite{yfallah:tvtsupra2016} scheme enables UEs to control their communication range and avoid excessive packet collisions based on channel busy percentage (CBP), a metric for probing the network utilization which will be discussed in Section III\textit{.B.}. SUPRA algorithm employs the concept of mapping CBP to radiation power with a single-step memory. It maintains CBP in the range of $[U_{\text{min}} \,\, U_{\text{max}}]$ by governing the radiation power as formulated in Equ. \ref{equ:SAEpwr}, in which the last set value of power, $P_k$, is used to calculate the new radiated power value, $P_k+1$.
$$P_{k+1}=P_k + \eta \times \Big( g(\text{CBP}) - P_k \Big)$$
\begin{equation} \label{equ:SAEpwr}
g(\text{CBP}) = \begin{cases}
P_{\text{max}},              & \text{CBP}<U_{\text{min}}\\ 
P_{\text{min}} +  \Big( \frac{U_{\text{max}} - \text{CBP}}{U_{\text{max}} - U_{\text{min}}} \Big) \\
\, \, \, \, \,\,\,\,\,\,\,\,\times (P_{\text{max}} - P_{\text{min}}),                           &  U_{\text{min}} \le \text{CBP} < U_{\text{max}}\\
P_{\text{min}},              & U_{\text{max}} \le \text{CBP}.
\end{cases}
\end{equation}
in which $\eta$ is the smoothing factor and $P_{\text{min}}$ and $P_{\text{max}}$ denote the allowed power boundaries.
\section{Implementation Details}
For the purpose of our study, we developed a link-level event-based network simulator in NS-3 environment, details of MAC and PHY layer implementation are presented in \cite{btoghi:vnc}. In this work we mount DCC in the application layer, on top of the C-V2X stack.
\subsection{Simulation Configurations}
 In order to simulate the air interface effects, e.g., channel pathloss and fading, the 3GPP community have proposed a set of log-distance channel models known as Winner+ \cite{winner+}. One minor contribution of this work is utilizing a realistic propagation channel model from our previous research in \cite{campVSC3phase1}. The Fowlerville channel model is derived from a large data set, collected during field trials on FTT-A Fowlerville Proving Ground by the Crash Avoidance Metrics Partnership (CAMP) consortium in collaboration with USDoT. This realistic channel model consists of both large and small-scale propagation effects. The anomalous null-pattern in figures \ref{fig:fig2}, \ref{fig:fig3}, \ref{fig:fig6}, and \ref{fig:fig7} is an artefact of this channel model \cite{ghayoor:emulator}.

We study a VANET consisting of moving vehicles in a $3.6 km$ straight highway with 12-lanes in two directions. Since vehicles closer to the end of the highway have fewer neighbors in the simulation, we observe what is called edge effect arising due to discontinuity in the edges. To avoid this, we extract our results only from the perspective of vehicles in the middle $1.2 km$ stretch. Vehicle speeds are determined based on the suggested values in \cite{3gpp:36885}. The initial $10 s$ of all simulations are disregarded to allow the network to stabilize. Other simulation assumptions and configurations are listed in Table \ref{table:configs}.
\begin{table}[t]
\centering
\caption{Simulation Parameters \& Configurations}
\begin{center}
\bgroup
\def\arraystretch{1.4}
\begin{tabular*}{0.37\textwidth}{@{\extracolsep{\fill} }  l r l r  }
\hline
\hline
Time $(T_{sim})$     & 120 s          & $\mathcal{B}$                & 25\\
Payload Size         & 190 B          & $\eta$                       & 0.5\\
MCS Index            & 5              & $P_{\text{min}}$             & 10 dBm\\
$\{T_1, T_2\}$       & $\{1, 100\}$   & $P_{\text{max}}$             & 23 dBm\\
Carrier Freq.        & 5860 MHz       & $U_{\text{min}}$             & 50\% \\
Bandwidth            & 10 MHz         &$U_{\text{max}}$              & 80\% \\
$\mathcal{P}_{\text{resel}}$    & 0.2               &$\text{ITT}^{\text{max}}$    & 600 ms\\
SLRRC                           & $\in$[5, 15]      &  $Th_{\text{SPS}}$          &-85 dBm\\
\hline
\end{tabular*}
\egroup
\label{table:configs}
\end{center}
\end{table}
\subsection{DCC-enabled C-V2X}
In Section II\textit{.B.}, we discussed the details of DCC suggested by \cite{sae:j2945}, however, this architecture is designed for DSRC communication and therefore needs modifications to be able to operate in C-V2X networks. Although 3GPP standard does not provide an explicit congestion control algorithm\cite{jgozalvez:vtm}, it suggests a framework based on channel occupancy \cite{3gpp:RAN:QC:DCC}. We examine this framework and propose a combined architecture for DCC-enabled C-V2X. In contrast with IEEE 802.11p, radio resources are divided in both time and frequency in the context of C-V2X. Hence, 3GPP revises the definition of CBP as follows:
\begin{itemize}
    \item CBP, measured by a UE at subframe number $n$, is defined as the portion of subchannels whose their measured S-RSSI exceeds a preset threshold value.
\end{itemize}

CBP is a metric to measure what percentage of the channel is being utilized by other nodes in the neighborhood of a UE. A second metric is defined to demonstrate what fragment of the channel is occupied by the UE itself; Channel-occupancy Ratio (CR) for a given UE, measured at subframe number $n$, is defined in Equ. \ref{equ:3gppCR}.
\begin{equation} \label{equ:3gppCR}
\begin{gathered}
\text{CR}(n) =  \frac {\sum\limits^{\tau_2}_{j=\tau_1} \sum\limits^{N_{\text{SubCH}}}_{i=1} x_{i,j}\theta_{i,j}}{\sum\limits^{\tau_2}_{j=\tau_1} \sum\limits^{N_{\text{SubCH}}}_{i=1} x_{i,j}} \\
\textit{s.t. \quad }\tau_2-\tau_1=1000ms, \quad n-\tau_1>  \left\lvert \tau_2-\tau_1 \right\lvert /2\\
\end{gathered}
\end{equation}
in which $N_{\text{SubCH}}$ is the number of subchannels per subframe (in our case $N_{\text{SubCH}}=2$). For the $i$th subchannel of subframe number $j$, the binary indicator, $x_{i,j}$, is set to 1 if that subchannel belongs to UE's resource pool. $\theta_{i,j}$ is the binary reservation indicator and $\theta_{i,j}=1$ indicates that subchannel is used or reserved by the UE (green sections in Fig. \ref{fig:fig1}).
 
The CBP measurement can be used to estimate the number of neighboring nodes in a UE's proximity, $N_{STA}$. The 3GPP standard suggests a mapping functional to estimate $N_{STA}$ using CBP and limit CR in order to maintain a preset CBP threshold. Equ. \ref{equ:3gppcc} formulates the described logic to derive $\text{CR}_{\text{limit}}$, the limit on the fraction of radio resources that a UE can utilize for its transmissions.
\begin{equation} \label{equ:3gppcc}
\textbf{if: } \text{CBP} > \text{CBP}_{\text{limit}}\,, \quad
\textbf{then: }  \text{CR}_{\text{limit}} = \frac{\text{CBP}_{\text{limit}}}{f^{-1}(\text{CBP})}\,.
\end{equation}
in which $\text{CBP}=f(N_{STA})$ is the mapping functional to estimate the number of neighboring vehicles and $\text{CBP}_{\text{limit}}$ is the preset CBP threshold.

We employ the same CBP definition and connect it to the Range Control algorithm of Equ. \ref{equ:SAEpwr}. However, our simulation results in Section IV show that the proposed method of mapping CBP to $N_{STA}$ is not effective in high-density scenarios. Hence, we directly count the number of unique vehicles within the vehicle's $100 m$ proximity, as described in \cite{sae:j2945}. This measure is then ported to the Rate Control algorithm in Equ. \ref{equ:SAEITT}.
\section{Analysis and Results}
In our previous work \cite{btoghi:vnc}, we utilized packet delivery ratio (PDR) and inter-packet gap (IPG) as key performance indicators and evaluated the baseline performance of C-V2X communication under high network load. PDR is defined as the number of received packets divided by the number of transmitted packets averaged over all node-pairs. IPG, measured between a UE-pair, is the average time duration between two consecutive packets that a UE receives from the other. It should be noted that DCC manipulates the transmission rate and range, thus, these metrics might not be sufficient and can be misleading in some situations.

To elaborate further, PDR does not represent a fair comparison between scenarios with different transmission rates, i.e., different numbers of total transmitted packets.
In order to provide an equal and meaningful comparison between baseline and DCC cases in different scenarios, we consider a third performance indicator to gauge both communication latency and reliability. We define Sidelink Throughput (SLT) as the data volume per time that a UE receives from a neighboring UE, averaged over all node-pairs. SLT, alongside with previously introduced metrics, is employed to evaluate the performance of DCC algorithm in C-V2X communication. 
\subsection{Baseline vs. DCC-enabled C-V2X}
\begin{figure}[t]
\includegraphics[width=.48\textwidth, left]{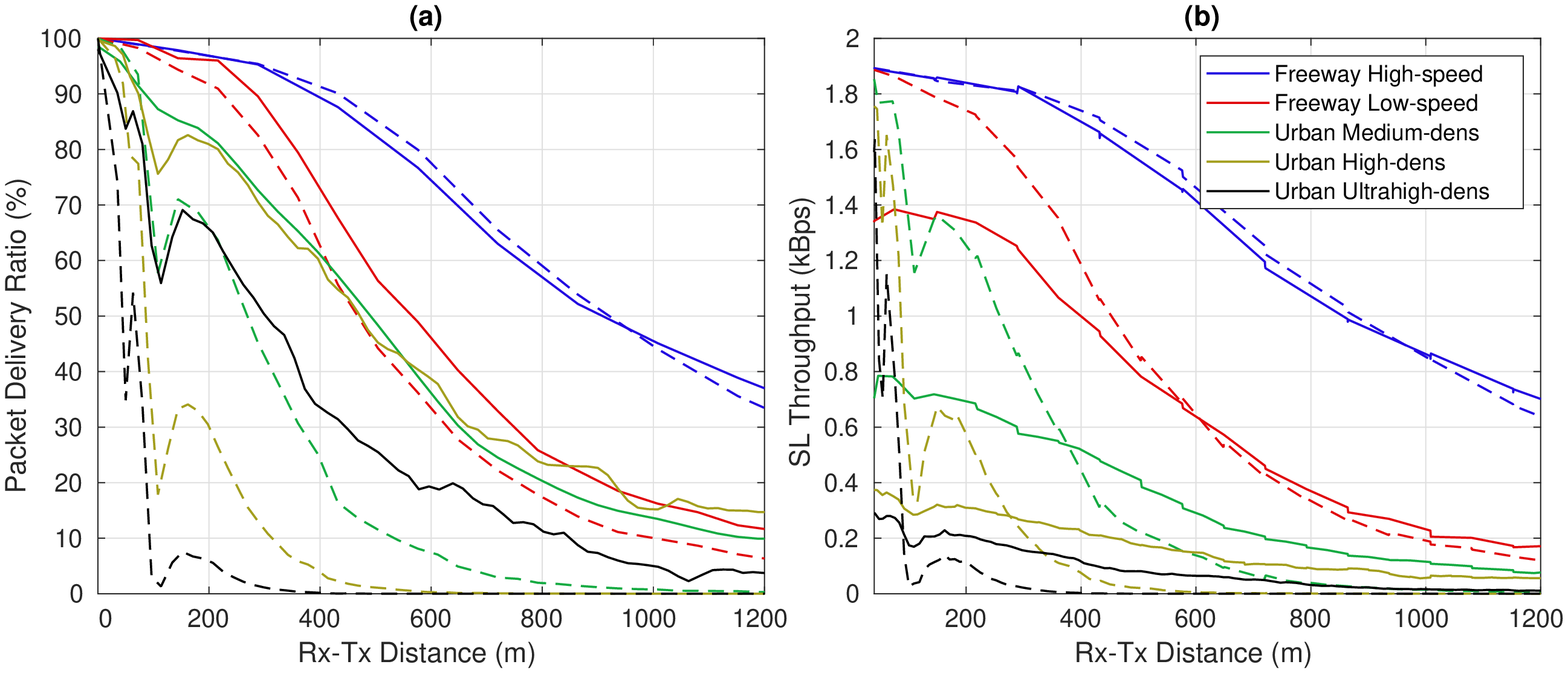}
\caption{Performance comparison of baseline (\textit{dashed}) and DCC-enabled (\textit{solid}) C-V2X in different density and mobility scenarios. Shown in terms of: \textit{(a)} packet delivery ratio, \textit{(b)} sidelink throughput.}
\label{fig:fig2}
\end{figure}
We conduct our comparative study
on 5 scenarios with different node densities and mobility models as suggested in \cite{3gpp:36885}. \textit{Freeway High-speed} scenario contains 300 vehicles with density of 7 Veh/(km.lane)\footnote{Vehicle per kilometer per lane}, cruising at 140km/h. \textit{Freeway Low-speed} scenario contains 600 vehicles with density of 14 Veh/(km.lane), cruising at 70km/h. \textit{Urban Medium-density}, \textit{High-density}, and \textit{Ultrahigh-density} scenarios accommodate 1200 (28 Veh/(km.lane)), 2400 (56 Veh/(km.lane)), and 4800 (111 Veh/(km.lane)) vehicles respectively, all moving at 15km/h.
\begin{figure}[t]
\centering
\includegraphics[width=.48\textwidth]{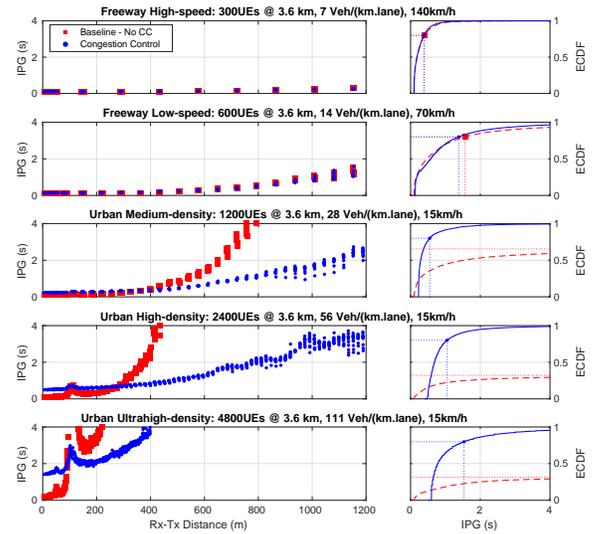}
\caption{Performance comparison of baseline (red squares) and DCC-enabled (blue circles) C-V2X in different density and mobility scenarios. \textit{Left}: Average IPG vs. distance, \textit{Right}: ECDF of IPG marked with 80th percentile.}
\label{fig:fig3}
\end{figure}
%
%
%
\subsubsection{Improvement in Network Capacity}
It is worth mentioning that in our simulation setup (Section III), every subframe fits two radio resources, as shown in Fig. \ref{fig:fig1}, and the selection window size is set to $100 ms$ which provides every UE with 200 radio resources. This is the maximum number of available radio resources to be allocated by UEs, thus, in the ideal case of a fully connected network and assuming perfect scheduling, the network cannot accommodate more than 200 UEs without collision. We refer to this limit as the \textit{saturation density} and all Urban and Freeway simulation scenarios are chosen to be over-saturated. Figures \ref{fig:fig2} and \ref{fig:fig3} illustrate the results for the above-mentioned scenarios. It can be observed that DCC mechanism does not intervene in the lowest density scenario, i.e., Freeway High-speed, due to the configured rate and range control parameters. The effect of DCC becomes noticeable in higher density scenarios, in which the baseline C-V2X degrades heavily as PDR drops to less than 10\% in relatively near distances and UEs lose communication (Fig. \ref{fig:fig2}a). This is also observable in Fig. \ref{fig:fig3}, where the baseline IPG grows rapidly in shorter distances.

Both SLT and IPG plots in figures \ref{fig:fig2}b and \ref{fig:fig3} demonstrate the design rationale behind Rate Control, that is sacrificing throughput in near ranges in order to maintain the communication reliability in farther distances. Fig. \ref{fig:fig2}b shows that SLT of the baseline mode rapidly decreases to less than 0.1 kBps in Urban scenarios, while DCC-enabled C-V2X reaches noticeably farther distances. Fig. \ref{fig:fig3} supports this claim and shows a rapid growth in average IPG for the baseline mode. Looking at the statistics of IPG, shown in Fig. \ref{fig:fig3} in the form of empirical cumulative distribution function (ECDF), is a more informative investigation. These ECDF plots demonstrate a wide difference between IPG distribution in baseline and DCC-enabled C-V2X. In the case of $\text{SLT}\xrightarrow{}0, \,\overline{\text{IPG}}\xrightarrow{}\infty$, we face the problem of blind-nodes which means there are vehicles that have not received any message from their region of interest (specified by $Th_{\text{SPS}}$) during the whole simulation time. CVS applications become absolutely incompetent in such cases, which can be avoided leveraging DCC. The impact of DCC becomes more striking in higher densities, e.g., DCC outperforms baseline in almost all distances in the case of Urban Ultrahigh-density scenario (Fig. \ref{fig:fig2}b).
\begin{figure}[b]
\centering
\includegraphics[width=.48\textwidth]{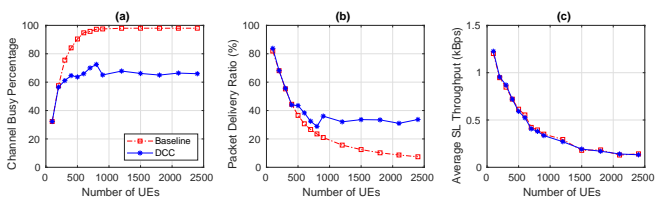}
\caption{Impact of DCC on network capacity.}
\label{fig:fig4}
\end{figure}
\begin{figure}[b]
\centering
\includegraphics[width=.48\textwidth]{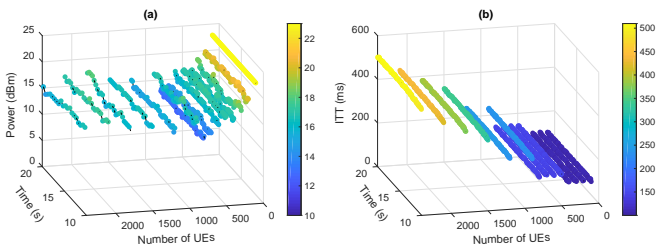}
\caption{Effect of node density on (a): Range Control, (b) Rate Control.}
\label{fig:fig5}
\end{figure}
Fig. \ref{fig:fig4} summarizes the comparison between baseline and DCC-enabled C-V2X. Fig. \ref{fig:fig4}b illustrates how DCC avoids communication degradation, in terms of PDR, compared to the baseline. Another noticeable point is, although Rate Control decreases the transmission rate in high density cases, Fig. \ref{fig:fig4}c shows that the average SLT has not dropped, compared to baseline. This is also consistent with Fig. \ref{fig:fig2}b as area-under-the-curve does not significantly change between baseline and DCC plots. All in all, we have observed that DCC strongly improves PDR, maintains the latency, extends the communication range, and avoids the blind-node situation without heavily sacrificing the average sidelink throughput.
\subsubsection{Analysis on Rate and Range Control}
As discussed in Section II\textit{.B.}, DCC probes the vehicle density for Rate Control and relies on a closed-loop feedback scheme to govern the Range Control. As a matter of fact, $N_{STA}$ grows linearly with the number of UEs and Equ. \ref{equ:SAEITT} also denotes a (piece-wise) linear relationship between ITT and $N_{STA}$. It is witnessed in Fig. \ref{fig:fig5}b, which demonstrates a set of ITT time-series for different number of UEs, that ITT keeps increasing to the maximum allowed value, $\text{ITT}^{\text{max}}$. However, Fig. \ref{fig:fig5}a does not show the same trend. The radiation power almost plateaus when the number of UEs exceeds 800. In order to further understand this phenomena, we need to study the relationship between CBP and vehicle density.

An ideal resource allocation should maximize the channel utilization and do not waste radio resources by under-utilizing the channel. Thus, in over-saturated densities (as defined before), CBP should ideally be around 100\%, which is in contrast with Fig. \ref{fig:fig4}a. This behavior is due to imperfections of SB-SPS mechanism which over-allocates a radio resource by multiple UEs while the channel is not yet fully-utilized, meaning that there are radio resources that are experiencing packet collisions and some others are empty. On the other hand, under heavy network loads, e.g, more than 800 UEs, the channel is fully utilized and CBP is saturated at ${\sim}100\%$. By adding more load to the network, CBP does not change while radio resources experience more number of collisions. This is an important observation which indicates Range/Power Control reaches its saturation point even in medium density scenarios. We suggest that this is due to the definition of CBP by 3GPP and employing more appropriate metrics can enhance the Range Control as also shown in \cite{bkang:cv2xPowerControl}. 
%
\subsection{Impact of Tunable Parameters}
Previous studies in \cite{bazzi:access, jgozalvez:tvt2018, btoghi:vnc} have investigated the effect of MAC and PHY layer parameters on the performance of baseline C-V2X. In this section, we evaluate the effect of DCC's tunable parameters on achievable performance gain by conducting an extensive set of simulations with different DCC and C-V2X MAC parameters under different network loads. Comparing the simulation results, we propose recommendations to potentially enhance the congestion control. A sub-set of the tested DCC schemes are chosen for the purpose of our discussion here (Table \ref{table:CCschemes}). It should be noted that, in this section, we vary the earlier discussed standard simulation setup to alternative configurations in order to assess the impact of different parameters. We perform our analysis in the Urban High-density traffic model. These test schemes are then compared to baseline C-V2X and standard DCC (Table \ref{table:configs}). Results are demonstrated in terms of PDR Gain and SLT gain, defined as the difference between baseline and DCC performance; thus PDR Gain is shown in percent and SLT gain has the dimension of data-rate.
\begin{table}[t]
\centering
\caption{Test Congestion Control Schemes}
\begin{center}
\bgroup
\def\arraystretch{1.4}
\begin{tabular*}{0.39\textwidth}{@{\extracolsep{\fill} } c|ccccc }
DCC Scheme  &$P_{\text{max}}$   &$P_{\text{min}}$   &$U_{\text{max}}$   &$U_{\text{min}}$   & $\mathcal{B}$\\
\hline
\hline
DCC \#1     & 23dBm   & 23dBm    & 80\%      & 50\%      & 25    \\
DCC \#2     & 23dBm   & 10dBm    & 50\%      & 30\%      & 25    \\
DCC \#3     & 23dBm   & 5dBm     & 50\%      & 30\%      & 25    \\
DCC \#4     & 23dBm   & 5dBm     & 50\%      & 30\%      & 35    \\
DCC \#5     & 23dBm   & 5dBm     & 50\%      & 30\%      & 45    \\
DCC \#6     & 23dBm   & 5dBm     & 50\%      & 30\%      & 55    \\
DCC \#7$^{\mathrm{a}}$     & 23dBm   & 0dBm     & 50\%      & 30\%      & 45    \\
\hline
\multicolumn{4}{l}{$^{\mathrm{a}} \mathcal{P}_{\text{resel}}=0.2$, SLRRC$\in$[1, 5].}\\
\end{tabular*}
\egroup
\label{table:CCschemes}
\end{center}
\end{table}
\subsubsection{Impact of Rate Control Parameters}
\begin{figure}[h]
\centering
\includegraphics[width=.48\textwidth]{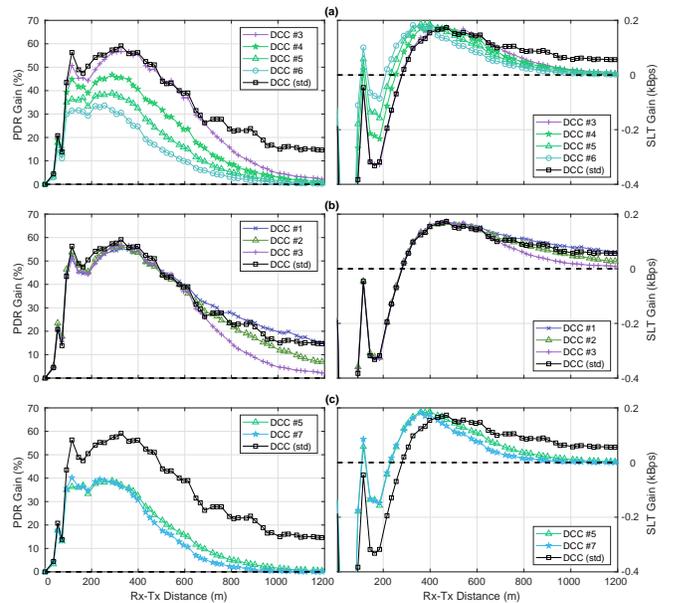}
\caption{Comparison between DCC schemes defined in Table \ref{table:CCschemes} and impact of (a) Rate Control, (b) Range Control, and (c) MAC layer parameters. DCC (std) is defined in Table \ref{table:configs}.}
\label{fig:fig6}
\end{figure}

Equ. \ref{equ:SAEITT} shows the linear mapping between the smoothed vehicle density, $N_{STA}^S$, and inter-transmission time, ITT. Slope of this function is determined by $\text{ITT}^{\text{max}}$ and Density Coefficient, $\mathcal{B}$. While the former is bounded to latency requirements of the CVS application, the latter can be used to tune the Rate Control mechanism. Fig. \ref{fig:fig6}a shows DCC schemes with identical Power Control configurations but less aggressive Rate Control (DCC \#3, \#4, \#5, and \#6), compared to the standard DCC. Results in this figure revisit the trade-off between SLT and PDR and imply that throughput in near distances can be improved with compromise on PDR by employing a less aggressive Rate Control Scheme. In the case of short-range safety applications, SLT loss (negative gain) in near ranges, can be mitigated by such tractable DCC schemes. We conclude that the choice of Rate Control scheme is conditioned on the CVS application and its corresponding critical situational awareness radius. Some applications such as Forward Collision Avoidance (FCA) give more importance to information from their near proximity and need a higher update rate in order to be able to precisely track the leading car, while other applications such as Green Light Optimized Speed Advisory (GLOSA) \cite{katsaros:glosa} need information from farther distances with much lower rate. Thus, the Rate Control can be optimized based on the safety importance of different CVS applications.
\subsubsection{Impact of Range Control Parameters}
The communication range is directly proportional to the radiated power, determined by Equ. \ref{equ:SAEpwr}. We examine more aggressive Range Control schemes (DCC \#2 and DCC \#3) with wider power range as well as Range Control deactivated (DCC \#1) scheme. Results shown in Fig. \ref{fig:fig6}b and Fig. \ref{fig:fig7} indicate that more aggressive Range Control schemes not only do not make an improvement but also downgrade the performance in far distances, due to the decreased link budget. A very important observation is related to the Range Control deactivated scheme (DCC \#1) which is, in fact, solely employing the Rate Control and performs almost identical to the standard DCC and even outperforms it in far distances. This matter indicates room for improvements on the current Range Control design and perhaps using other feedback probes instead of the current form of CBP, as discussed in Section IV\textit{.A.2.}
\subsubsection{Interoperability with C-V2X MAC}
\begin{figure}[t]
\centering
\includegraphics[width=.48\textwidth]{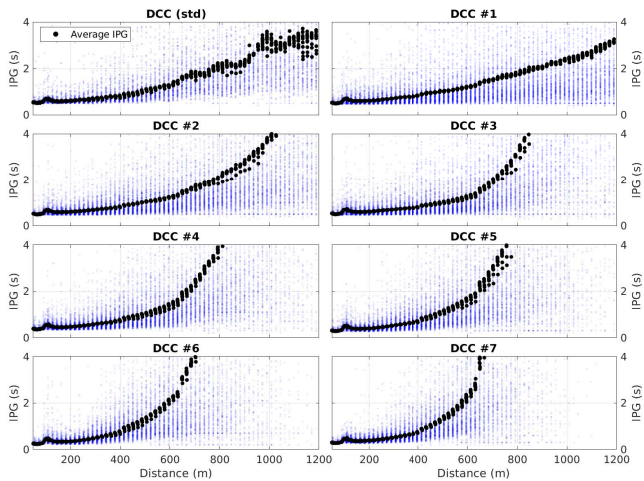}
\caption{IPG comparison of DCC schemes defined in Table \ref{table:CCschemes}. Distribution (\textit{blue}) and mean values (\textit{black}) are illustrated. DCC (std) is defined in Table \ref{table:configs}.}
\label{fig:fig7}
\end{figure}

Previous studies in 3GPP TSG-RAN technical reports have expressed concerns about possible conflicts between SB-SPS operation and variable transmission rate in the application layer, as they have different characteristic times and are not necessarily time-synchronized \cite{3gpp:RAN:beaconperiod}. In order to study this issue, we should revisit the SB-SPS algorithm. When a UE selects a radio resource, it periodically repeats it for a random number of transmissions, SLRRC. Expected value of the period that a UE occupies a radio resource before changing it, is a function of $\mathcal{P}_{\text{resel}}$ and SLRRC. Figures \ref{fig:fig6} and \ref{fig:fig7} illustrate a comparison for less inert DCC schemes. No significant performance change has observed via varying the MAC parameters in DCC \#5 and DCC \#7. We infer that the resource re-selection procedure does not become a bottle-neck when application layer varies the transmission rate. This observation is also aligned with the more detailed investigation in \cite{btoghi:cavs2019} which shows the time behavior of congestion control parameters over time.

%
%
\section{Concluding Remarks}
Recently, Cellular Vehicle-to-everything (C-V2X) communication has received significant interest as an alternative for Dedicated Short Range Communication (DSRC). Previous studies have shown that DSRC requires a sophisticated congestion control mechanism to combat performance degradation in high density environments. Hence, distributed congestion control (DCC) algorithm was standardized by SAE for DSRC to improve the performance under high network loads. Early investigations of C-V2X and our own study so far indicates that C-V2X also requires such a mechanism. In this work, we proposed a combined DCC-enabled design and evaluated the performance of transmission rate and range control components of DCC. Our analyses demonstrate that rate control has more significant impact on performance, compared to range control. While this work presents an implementation of DCC on C-V2X, it also indicates that in order to obtain optimal performance in congested environments, further exploration is needed to arrive at a more efficient range control scheme and appropriate settings for tunable parameters of C-V2X.
\balance
\bibliography{refs.bib}{}
\bibliographystyle{unsrt}
\end{document}